\documentstyle[11pt,moriond,epsfig]{article}

\def\be{\begin{equation}}
\def\ee{\end{equation}}
\def\bea{\begin{eqnarray}}
\def\eea{\end{eqnarray}}

\def\braket#1#2{\langle #1 |#2 \rangle}
\def\cm{{\cal M}}

\begin{document}


\vspace*{4cm}
\title{OVERVIEW OF NNLO QCD CORRECTIONS}

\author{ E.W.N. GLOVER }

\address{Department of Physics, University of Durham, South Road,\\
Durham DH1 3LE, England}

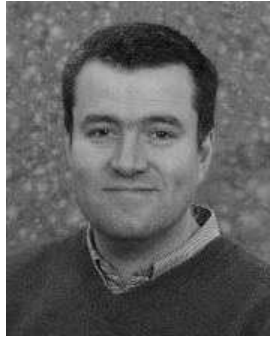
\maketitle\abstracts{
We discuss the motivation for making predictions for jet cross sections at
next-to-next-to-leading order.   We describe the theoretical ingredients needed
for such a calculation and briefly review the progress in the field.}


In the last decade there has been enormous progress in using perturbative QCD
to predict and describe events containing jets. At the simplest lowest-order
(LO) level,  each jet is the footprint of a hard, well separated parton
produced in the event. Although the predicted rate is sensitive to the choices
of renormalization and factorization scales, qualitative comparisons of data
and theory are generally very good. A more quantitative description is
achieved by improving the theoretical prediction to next-to-leading order
(NLO). This has the effect of reducing the dependence on the unphysical
renormalization and factorization scales so that the normalization is more
certain. Furthermore, the sensitivity to the details of the jet finding
algorithm and the size of the jet is increased since now two partons may be
combined to form the jet.  However, for the most basic jet production
processes such as $p\bar p \to $~jet + X, $p\bar p \to $V + jet + X, $e^+e^-
\to 3$~jets or  $ep \to e+(2+1)$~jets, the experimental accuracy is such that
even more precise theoretical predictions are required.   In this talk, we
review the recent progress made towards predicting jet cross sections at 
next-to-next-to-leading order (NNLO). 

The addition of NNLO effects gives significant improvements over an NLO
estimate. First, the dependence on the unphysical scales is significantly
reduced. At NLO we find a reliable estimate of the cross section, while
NNLO calculations yield a reliable estimate of the uncertainty in the cross
section. For example, if we consider the single jet inclusive cross section 
for jets with $E_T = 100$~GeV and $0.1 < |\eta| < 0.7$ at $\sqrt{s} =
1800$~GeV, the renormalisation scale dependence due to variations of a factor
of two about $\mu_R = E_T$ is reduced from 20\% to 9\% to 1\% as we move from
LO to NLO to NNLO.  The NNLO estimate is obtained by including the
renormalisation group predictable parts of the $\alpha_s^4$ contribution and
assuming that the (presently unknown) genuine NNLO contribution is zero.  Other
ingredients such as the factorisation scale and choice of parton density
functions are kept fixed.  Second, the sensitivity to the jet algorithm is
further enhanced with up to three partons combining to form the jet.
Radiation outside the jet is better described and more of the parton shower is
explicitly reconstructed.

There are several ingredients necessary for NNLO calculations of, for example, 
 jet production in
hadron colliders.  
From the matrix element point of view we need 
\begin{itemize}
\item  the interference of the tree and two-loop amplitudes for the  two~parton
final state,
\item  the square of the one-loop amplitude  for the  two~parton final state,
\item  the interference of the tree and one-loop amplitudes for the three~parton
final state,
\item  the square of the tree amplitude for the four~parton final state.
\end{itemize}
Each of these contributions is infrared divergent and a systematic 
procedure for analytically cancelling the infrared divergences
between the tree-level $2 \to 4$, the one-loop $2 \to 3$ and the $2\to 2$
processes needs to be established.
In addition, for physical cross sections, the parton density functions are
needed at NNLO which in turn requires knowledge of the three-loop splitting
functions as well as a global fit to other observables (such as deep inelastic
scattering and Drell-Yan) computed to NNLO.  
Finally, and most importantly for phenomenological applications, a numerical
implementation of the various contributions must be developed. 

\subsection*{Matrix elements}

Techniques for computing multiparticle tree amplitudes for  $2\to 4$ processes,
and the associated crossed processes, are well understood. For example, the
helicity amplitudes for the six gluon $gg\to gggg$, four gluon-two quark 
$\bar{q}q\to gggg$, two gluon-four quark $\bar{q}q\to\bar{q}^\prime q^\prime
gg$ and six quark $\bar{q}q\to\bar{q}^\prime q^\prime\bar{q}^{\prime\prime}
q^{\prime\prime}$ have been computed in Refs.~\cite{6g}. 
Similarly, amplitudes for the one-loop $2 \to 3$ parton sub-processes $gg\to
ggg$, $\bar{q}q\to ggg$, $\bar{q}q\to\bar{q}^\prime q^\prime g$, and processes
related to these by crossing symmetry, are also known and are available
in~\cite{5g} respectively.   However, the evaluation of the two-loop
$2\to 2$ contributions for QCD processes has been a  challenge for the past few
years, mainly due to a lack of knowledge about planar and crossed double box
integrals that arise at this level.  

In the massless parton limit and in dimensional regularisation, analytic
expressions   for these basic scalar integrals  have now been provided by
Smirnov~\cite{planarA} and Tausk~\cite{nonplanarA}  as series  in
$\epsilon=(4-D)/2$, where $D$ is the space-time dimension, together with
algorithms for reducing tensor integral to a basis set of known scalar (master)
integrals~\cite{planarB}.  This makes the calculation of the
two-loop amplitudes for $2 \to 2$ QCD scattering processes possible.


The general strategy is to rewrite   the
interference of two-loop with tree graphs in terms of a limited number of
master loop integrals, which can then be expanded as a series in $\epsilon =
(4-D)/2$, so that 
\begin{equation}
\label{eq:twoloop}
\braket{\cm^{(0)}}{\cm^{(2)}}+\braket{\cm^{(2)}}{\cm^{(0)}} =
\sum_{i=1}^4\frac{X_i}{\epsilon^i} + X_0. 
\end{equation}  
After renormalisation, the infrared singularities indicated by poles in the
dimensional regulator $\epsilon$ correspond to soft and/or collinear virtual
emissions.   The structure of these singularities can be predicted using the
general factorisation formula of Catani~\cite{catani} and serves as a powerful
check on the calculation. Following on from the pioneering work of Bern,  Dixon
and Ghinculov~\cite{BDG} who calculated the two-loop contribution to the QED  processes $e^+e^- \to \mu^+\mu^-$ and
$e^+e^- \to e^-e^+$, the ${\cal O}({\alpha_s^4})$ contributions arising from
the interference of two-loop and tree-level graphs for the QCD processes of
quark-quark, quark-gluon and gluon-gluon
scattering~\cite{gggg} have now been computed.   Similar techniques can be
applied to other two-loop processes with massless internal particles and
on-shell external legs such as $gg \to \gamma \gamma$ or $ q \bar q \to \gamma
\gamma$. 

Differential equation techniques have been developed for computing two-loop
master integrals with off-shell legs~\cite{diffeq} and
all of the planar and non-planar~\cite{offplanar} master
integrals are now known with one off-shell leg.   Such integrals are vital for 
basic scattering processes such as $Z \to q \bar q g$ and $ep \to (2+1)$~jets.

\subsection*{Infrared cancellation}

The infrared singularities present in the vitrual contributions
(Eq.~\ref{eq:twoloop}) must cancel against the contributions from the one-loop
$2 \to 3$ processes when one particle is unresolved and the contribution from
the tree-level $2 \to 4$ processes when two particles are unresolved. 
Unresolved particles are either soft or collinear with one of the other partons
in the event and both of these configurations have the appearance of a $ 2 \to
2$ scattering.      A systematic procedure for analytically carrying through
the cancellation (as well as providing a series of counter terms that remove
the divergence from the radiative contributions) has not yet been
established.   However, single and double unresolved limits of the matrix
elements are well known and this may be a tractable problem.   For example, in
the limit where three particles are simultaneously collinear, the (colour
ordered) tree amplitude undergoes a factorisation of the form~\cite{tc}
$$
\cm^{(0)}(\ldots,a,b,c,\ldots) \to 
P_{abc \to d} \cm^{(0)}(\ldots,d,\ldots)
$$
In other words, we find the tree level amplitude with 
particles $a$, $b$ and $c$ replaced with a single particle  $d$ multiplied by a single (universal) factor describing the
splitting $d \to a + b + c$.   All of the divergences associated with 
the  collinear singularities are contained inside the splitting function 
$P_{abc \to d}$ and, when integrated over the triple collinear phase space 
yield
singular contributions up to $1/\epsilon^3$.  Similar factorisation formulae
apply in the double soft~\cite{ds} and soft-collinear~\cite{tc} limits.

The one-loop amplitudes are also singular in the single unresolved limit -
either due to two particles becoming collinear or a soft gluon.   Once again
there is a factorisation~\cite{sone,cone} 
that yields a tree unresolved factor multiplied by a
one-loop amplitude with one fewer external particle plus a loop unresolved
factor multiplied by a tree amplitude.


Many of the analytic phase space integrations for the  double unresolved and
single unresolved loop contributions have already been studied in the context of
$e^+e^- \to $~photon + jet
at ${\cal O}(\alpha \alpha_s)$~\cite{aude} and Higgs production in hadron
colliders~\cite{grazzini}. 

\subsection*{Parton density Functions}

A further complication is due to initial state radiation. Factorization of
the collinear singularities from the incoming partons requires the evolution
of the parton density functions to be known to an accuracy matching that of
the hard
scattering matrix element.  This entails knowledge of the three-loop
splitting functions.  At three-loop order, the even moments of the splitting
functions are known for the flavour singlet and non-singlet structure
functions $F_2$ and $F_L$ up to $N=12$ while the odd moments up to $N=13$ are
known for $F_3$~\cite{moms1}. The numerically small $N_F^2$ non-singlet
contribution is also known~\cite{Gra1}. Van Neerven and Vogt have provided
accurate parameterisations of the splitting functions in
$x$-space~\cite{NV} which are now starting to be implemented in the
global analyses~\cite{MRS}.

\subsection*{Outlook}

While much work remains to be done, the current rate of progress suggests that 
numerical estimates of jet cross
sections at NNLO may become available in the next couple of years.    The
theoretical uncertainty at NNLO should be significantly reduced compared to NLO
estimates enabling more stringent tests of QCD at short distances.


\end{document}